\documentstyle[12pt,psfig,epsf]{article}
\newcommand{\be}{\begin{equation}}
\newcommand{\ee}{\end{equation}}

\newcommand{\beqa}{\begin{eqnarray}}
\newcommand{\eeqa}{\end{eqnarray}}

\newcommand{\eqref}[1]{(\ref{#1})}

\newcommand{\bt}{\beta}

\def\boxit#1{\vbox{\hrule\hbox{\vrule\kern8pt
\vbox{\hbox{\kern8pt}\hbox{\vbox{#1}}\hbox{\kern8pt}}
\kern8pt\vrule}\hrule}}
\def\mathboxit#1{\vbox{\hrule\hbox{\vrule\kern8pt\vbox{\kern8pt
\hbox{$\displaystyle #1$}\kern8pt}\kern8pt\vrule}\hrule}}

\def\IB{\relax\hbox{$\inbar\kern-.3em{\rm B}$}}
\def\IC{\relax\hbox{$\inbar\kern-.3em{\rm C}$}}
\def\ID{\relax\hbox{$\inbar\kern-.3em{\rm D}$}}
\def\IE{\relax\hbox{$\inbar\kern-.3em{\rm E}$}}
\def\IF{\relax\hbox{$\inbar\kern-.3em{\rm F}$}}
\def\IG{\relax\hbox{$\inbar\kern-.3em{\rm G}$}}
\def\IGa{\relax\hbox{${\rm I}\kern-.18em\Gamma$}}
\def\IH{\relax{\rm I\kern-.18em H}}
\def\IK{\relax{\rm I\kern-.18em K}}
\def\IL{\relax{\rm I\kern-.18em L}}
\def\IP{\relax{\rm I\kern-.18em P}}
\def\IR{\relax{\rm I\kern-.18em R}}
\def\IZ{\relax\ifmmode\mathchoice
{\hbox{\cmss Z\kern-.4em Z}}{\hbox{\cmss Z\kern-.4em Z}}
{\lower.9pt\hbox{\cmsss Z\kern-.4em Z}} {\lower1.2pt\hbox{\cmsss
Z\kern-.4em Z}}\else{\cmss Z\kern-.4em Z}\fi}

\def\II{\relax{\rm I\kern-.18em I}}

\begin{document}

\hfill  NRCPS-HE-01-30

\hfill  NTUA-9/01

\vspace{24pt}
\begin{center}
{\large \bf Three-dimensional gonihedric spin system}

\vspace{24pt}

{\sl G.Koutsoumbas}\\
Physics Department, National Technical University, \\
Zografou Campus, 15780 Athens, Hellenic Republic\\
{\tt email:kutsubas@central.ntua.gr}
\vspace{1cm}

{\sl G.K.Savvidy}

National Research Center Demokritos,\\
Ag. Paraskevi, GR-15310 Athens, Hellenic Republic\\
{\tt email:savvidy@mail.demokritos.gr}
\end{center}
\vspace{60pt}

\centerline{{\bf Abstract}}

\vspace{12pt} \noindent We perform Monte Carlo simulations of a
three-dimensional spin system with a Hamiltonian which contains
only four-spin interaction term. This system describes
random surfaces with extrinsic curvature - gonihedric action.
We study the anisotropic model when the coupling constants $\beta_S$ for
the space-like plaquettes and $\beta_T$ for the transverse-like
plaquettes are different. In the two limits $\beta_S=0$ and
$\beta_T=0$ the system has been solved exactly and the main interest
is to see what happens when we move away from these points towards
the isotropic point, where we recover the original model.
We find that the phase transition is of first order
for $\beta_T = \beta_S \approx 0.25,$ while away from this point
it becomes weaker and eventually turns to a crossover.
The conclusion which can be drown from this result is that the exact
solution at the point $\beta_S =0$ in terms of 2d-Ising model should be
considered as a good zero order approximation in the
description of the system also at the isotropic point $\beta_S =\beta_T$
and clearly confirms the earlier findings that at the isotropic
point the original model shows a first order phase transition.


\newpage

\pagestyle{plain}
\section{Introduction}


In this article we  shall consider a model of two-dimensional
random surfaces embedded into a Euclidean lattice $Z^3,$ where a
closed surface is associated with a collection of plaquettes. The
surfaces may have self-intersections in the form of four
plaquettes intersecting on a link. Various  models of  random
surfaces built out of plaquettes have been considered in the
literature \cite{weingarten}. The gas of random surfaces defined
in \cite{weg} corresponds to the partition function with Boltzmann
weights proportional to the total number of plaquettes. In this
article we shall consider the so-called gonihedric model  with
extrinsic curvature action \cite{sav1,sav}.  The gonihedric model
of random surfaces corresponds to a statistical system with weights
proportional to the total number of non-flat edges $n_2$ of the
surface \cite{sav1}. The weights associated with
self-intersections are proportional to $k n_4$ where $n_4$ is the
number of edges with four intersecting plaquettes, and $k$ is the
self-intersection coupling constant \cite{sav1,sav}. The partition
function is a sum over two-dimensional surfaces of the type
described above, embedded in a three-dimensional lattice: \be
Z(\beta) = \sum_{\{surfaces~M\}} e^{-\beta~\epsilon(M)},
\label{partfan} \ee where $\epsilon(M)=n_2 + 4 k n_4$ is the
energy of the surface $M$.

In three dimensions the equivalent spin Hamiltonian is equal to
\cite{sav}
 \be H_{gonihedric}^{3d}=- 2k \sum_{\vec{r},\vec{\alpha}}
\sigma_{\vec{r}} \sigma_{\vec{r}+\vec{\alpha}} + \frac{k}{2}
\sum_{\vec{r},\vec{\alpha},\vec{\beta}} \sigma_{\vec{r}}
\sigma_{\vec{r}+\vec{\alpha} +\vec{\beta}} -  \frac{1-k}{2}
\sum_{\vec{r},\vec{\alpha},\vec{\beta}} \sigma_{\vec{r}}
\sigma_{\vec{r}+\vec{\alpha}}
\sigma_{\vec{r}+\vec{\alpha}+\vec{\beta}}
\sigma_{\vec{r}+\vec{\beta}}, \label{hamil} \ee and it is an
alternative  model to the $3D$ Ising system \cite{weg}
$$
H_{Ising}^{3d}= -  \sum_{\vec{r},\vec{\alpha}} \sigma_{\vec{r}}
\sigma_{\vec{r}+\vec{\alpha}}.
$$
The degeneracy of the vacuum state depends on
self-intersection coupling constant $k$ \cite{pav}. If $k \neq 0$,
the degeneracy of the vacuum state is equal to $3\cdot 2^N$ for
the lattice of size $N^3,$ while it equals $2^{3N}$ when $k=0$.
The last case is a sort of supersymmetric point in the space of
gonihedric Hamiltonians \cite{pav}
\be
H_{gonihedric}^{k=0}=-\frac{1}{2}\sum_{\vec{r},\vec{\alpha},\vec{\beta}}
\sigma_{\vec{r}} \sigma_{\vec{r}+\vec{\alpha}}
\sigma_{\vec{r}+\vec{\alpha}+\vec{\beta}}
\sigma_{\vec{r}+\vec{\beta}}. \label{k=0case}\ee This enhanced
symmetry allows the construction of the dual Hamiltonian which has the
form \cite{pav} \be \label{dual} H^{k=0}_{dual} = -\sum_{\xi}
\left[ R^{\chi}(\xi) \cdot R^{\chi}(\xi + \chi) +R^{\eta}(\xi) \cdot
R^{\eta}(\xi + \eta)+ R^{\varsigma}(\xi) \cdot R^{\varsigma}(\xi +
\varsigma) \right], \ee where $\chi$,~$\eta$~and $\varsigma$ are unit
vectors in the orthogonal directions of the dual lattice and
$R^{\chi}$, $R^{\eta}$ and $R^{\varsigma}$ are one-dimensional
irreducible representations of the group $Z_{2} \times Z_{2}$.

\begin{figure}
\centerline{\hbox{\psfig{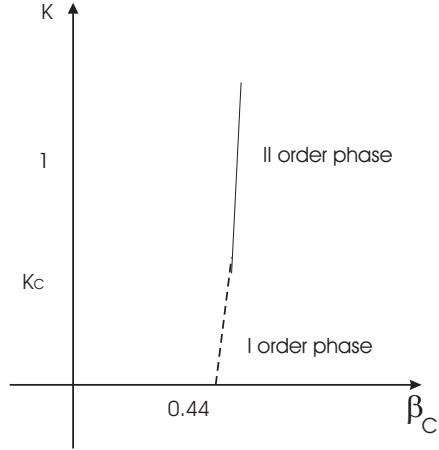}}}
\caption[fig1]{Phase diagram for different values of intersection
coupling constant $k$.}
\label{fig1}
\end{figure}

To study statistical and scaling properties of the system one can
directly simulate surfaces by gluing together plaquettes with the
corresponding weight $exp(-\beta (n_2 + 4 k n_4))$ or (much
easier) to study the equivalent spin system (\ref{hamil}).
The first Monte Carlo simulations
\cite{bath,des,koutsoumbas,cappi} demonstrate (see Figure \ref{fig1})
that the gonihedric system
with intersection coupling constant greater than $k_c \approx 0.5$
(including $k=1),$ undergoes a second order phase transition at
$\beta_{c} \approx 0.44$ and that the critical indices are
different from those of the 3D Ising model.
Thus they are in
different classes of universality. On the contrary, the system
shows a first order phase transition for $k < k_c,$
including the ``supersymmetric" point $k=0$.

Essential progress in our understanding of the physical behavior of the
system has been achieved by means of the transfer matrix
approach \cite{pav,transferma}.
The corresponding transfer matrix can be constructed for all
values of the intersection coupling constant $k$ \cite{george} and
describes the propagation of the closed loops in the time direction.
In this article we shall consider only the $k=0$ case. The
corresponding transfer matrix has the form \cite{pav} \be
K(Q_{1},Q_{2}) = \exp \{-\beta ~[ k(Q_{1}) + 2 l(Q_{1}
\bigtriangleup Q_{2}) + k(Q_{2})]~ \}, \label{tranmat} \ee where
$Q_{1}$ and $Q_{2}$ are closed polygon-loops on a two-dimensional
lattice, $k(Q)$ is the curvature and $l(Q)$ is the length of the
polygon-loop $Q$ \footnote{We shall use the word ``loop" for the
``polygon-loop".}. This transfer matrix describes the propagation
of the initial loop $Q_{1}$ to the final loop $Q_{2}$ and    
corresponds exactly to the Hamiltonian (\ref{k=0case}),(\ref{dual}).
Thus in order to study the critical behavior  of the
$k=0$ system  (\ref{k=0case}),(\ref{dual}),
(\ref{tranmat}) one should find the spectrum of the transfer matrix
(\ref{tranmat}). Generally speaking all three dimensional 
problems of statistical
mechanics are extremely complicated and in our case the exact
solution still remains out of reach.
However the breakthrough comes from the exact solution of a closed
system when the transfer matrix depends only on the 
symmetric difference of initial and final loops $Q_{1}
\bigtriangleup  Q_{2}$ \cite{transferma,george}
\be \tilde{K}(Q_{1},Q_{2}) = \exp \{-2
\beta  l(Q_{1} \bigtriangleup Q_{2}) ~ \}. \label{subtranmat}
\ee
The spectrum of the last system has been evaluated analytically in terms
of correlation functions of the 2D Ising model. This is a nontrivial
example of exactly solvable system in three dimensions.
In particular the largest eigenvalue $\Lambda_{0}$
is exactly equal to the
partition function of the 2d Ising model and  the free energy
of our 3d system is therefore equal to the free energy of the 2d Ising model.
This result nicely explains why the critical temperature of the
three-dimensional gonihedric system is so close to the critical
temperature of the two-dimensional Ising model  $2\beta_{c} =
\frac{1}{2} ln(1+\sqrt{2}) \approx 0.44$.
The next to the largest eigenvalue $\Lambda_{1}$ of the
3d system (\ref{subtranmat}) coincides with internal energy
$u(\beta)$ of the two-dimensional Ising model
\be
{\Lambda_{1} \over
\Lambda_{0} } ~=~ \langle \sigma_{\vec{r}} ~
\sigma_{\vec{r} + \hat{1}} \rangle ~=~-u(\beta)
\ee
and the correlation length $\xi(\beta),$ defined
through the ratio of eigenvalues $\Lambda_{1}/\Lambda_{0},$
is equal to:
$$
\xi(\beta) = {1\over -ln(\Lambda_{1}/\Lambda_{0})} =
{1\over -ln(-u(\beta))} .
$$
If the internal energy  $u(\beta)$ approaches $1,$ 
the correlation length tends to infinity and signals a  
second order phase transition in the system.
But the internal energy of the 2d-Ising model drastically
increases at the critical point
$2\beta_{c} =\frac{1}{2} ln(1+\sqrt{2})$ without reaching the value 1 !
Thus, in three dimensions, we have the extraordinary situation
that the specific heat has the logarithmic singularity of
the 2d-Ising model, but the
correlation length remains finite. The conclusion is that
the system undergoes a weak first order phase transition
rather than a second order phase transition.
The Hamiltonian which corresponds to  the transfer matrix
(\ref{subtranmat}) has been found in \cite{george} and is equal
to \be \tilde{H}_{Q_{1} \bigtriangleup  Q_{2}} =
\sum_{E_{x},E_{y}}\sigma\sigma\sigma\sigma ,\label{subtranmat1}
\ee where a summation is only over $vertical$ plaquettes,
the interactions take place only on the vertical planes
$E_x$ and $E_y$ and the ``horizontal" interactions have been
switched off.

The above consideration poses the following interesting question:
let us consider the system
\be
H_{anisot} = \beta_S \sum_{E_{z}}\sigma\sigma\sigma\sigma +\beta_T
\sum_{E_{x},E_{y}}\sigma\sigma\sigma\sigma, \label{ba}\ee
which has anisotropic coupling constants for vertical
and horizontal plaquettes. We have seen that
when $\beta_S =0$ the system reduces to the system
(\ref{subtranmat}),(\ref{subtranmat1}) and as we just explained 
it has been solved
in \cite{transferma,george}. When $\beta_T =0$ it factors into identical
two dimensional plane systems solved in \cite{pav} and it is always
in the disordered phase. Finally when $\beta_S = \beta_T$ we arrive at our original
$k=0$ system (\ref{k=0case}),(\ref{dual}),(\ref{tranmat}).
Thus we know the behavior of the system at both $\beta_S =0$ and
$\beta_T =0$, but we still don't know the analytical solution at the
isotropic point $\beta_S =\beta_T$. The understanding of
the phase structure of the anisotropic system (\ref{ba}) on
$(\beta_S,\beta_T)$ plane by means of Monte Carlo simulations can
drastically clarify the situation. Indeed, the important
question to which we would like to find
an answer  is whether or not there is any
dramatic changes in the behavior of the system when we move out of
solvable point $\beta_S =0$ to the isotropic point $\beta_S =\beta_T,$
where a first order phase transition has been observed
in \cite{des,cappi}.
Our Monte Carlo simulations show that there are no changes in the
behavior of the system
as it moves from $\beta_S =0$ to the isotropic point $\beta_S =\beta_T$.
The conclusion which can be drawn from this result is that the exact
solution at the point $\beta_S =0$ in terms of 2d-Ising model should be
considered as a good zero order approximation in the
description of the system also at the isotropic point $\beta_S =\beta_T$
if one consider a perturbation with the coupling constant $\beta_S/\beta_T$.
This should be checked by further analytical consideration.

\section{The lattice model}

Thus the lattice action (\ref{ba}) may be written in the form
$$ S \equiv \beta_S \sum P_{xy}+\beta_T \sum (P_{xz}+P_{yz}),$$

$$ P_{xy}(\vec{r}) \equiv 1-\sigma(\vec{r}) \sigma(\vec{r}+\hat{x})
\sigma(\vec{r}+\hat{x}+\hat{y}) \sigma(\vec{r}+\hat{y}),$$

$$ P_{xz}(\vec{r}) \equiv 1-\sigma(\vec{r}) \sigma(\vec{r}+\hat{x})
\sigma(\vec{r}+\hat{x}+\hat{z}) \sigma(\vec{r}+\hat{z}),$$

$$ P_{yz}(\vec{r}) \equiv 1-\sigma(\vec{r}) \sigma(\vec{r}+\hat{y})
\sigma(\vec{r}+\hat{y}+\hat{z}) \sigma(\vec{r}+\hat{z}).$$
Our goal is to find the phase diagram in the extended $\beta_S-\beta_T$ plane.
For that we shall calculate in the sequel the mean values of the action $S,$ the
space-like plaquette $P_S \equiv P_{xy}$ and the transverse-like plaquette
$P_T \equiv \frac{P_{xz}+P_{yz}}{2}.$
These quantities serve as order parameters
which help us identify the various phases.

A first attempt towards the determination of the phase diagram is through the
mean field approximation. One considers the free energy in the mean field
approximation, which (up to additive constants) is given by the expression:
$$F(x) = -(\beta_S+2 \beta_T) [u^\prime(x)]^4-u(x)+x u^\prime(x).$$
The function $u(x)$ is defined through the
relation $$\exp[u(x)] \equiv e^x+e^{-x} = 2 \cosh(x) \rightarrow u(x) =
\log(2)+\log[\cosh(x)] \rightarrow u^\prime(x)=\tanh(x).$$
We observe that $F(x)$ depends on the combination $B \equiv
\beta_S+2 \beta_T.$ The free energy has always a local minimum at $x=0.$
For small $B$ this is also the global minimum. As $B$ increases,
a second minimum shows up, which eventually wins and becomes the global
minimum at $B =0.688.$ Thus the phase transition line is given by:
$$\beta_T = \frac{0.688-\beta_S}{2}.$$ More accurately it is the
segment of this line which corresponds to positive values of $\beta_S$ and
$\beta_T.$ For $\beta_S=0$ we predict a critical value 0.344 for 
$\beta_T,$ while as $\beta_S$ increases the critical value decreases and
finally becomes zero at $\beta_S=0.688.$

\begin{figure}
\begin{tabular}{cc}
\psfig{figure=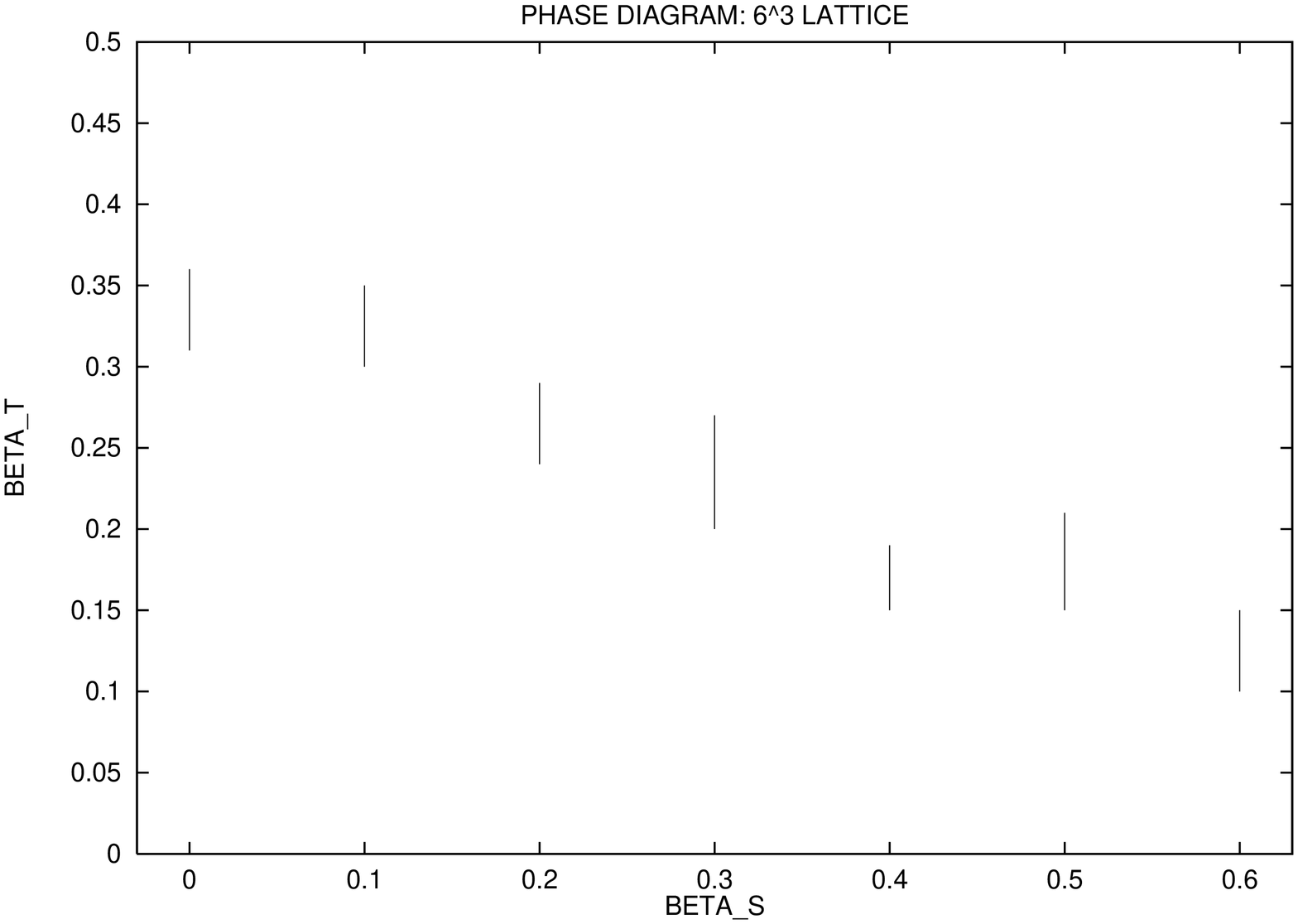,height=4cm,angle=-90}&
\psfig{figure=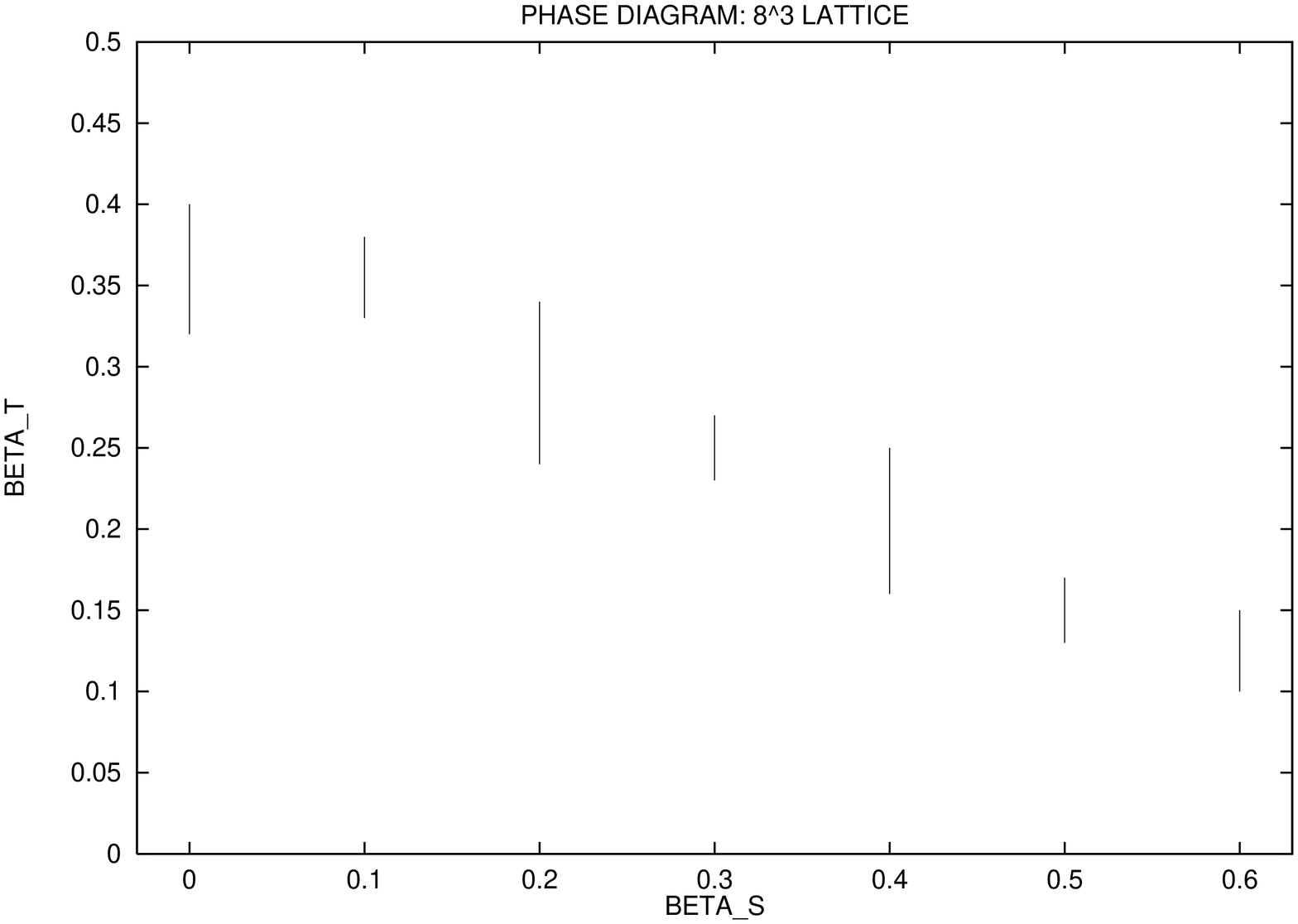,height=4cm,angle=-90}\\
\psfig{figure=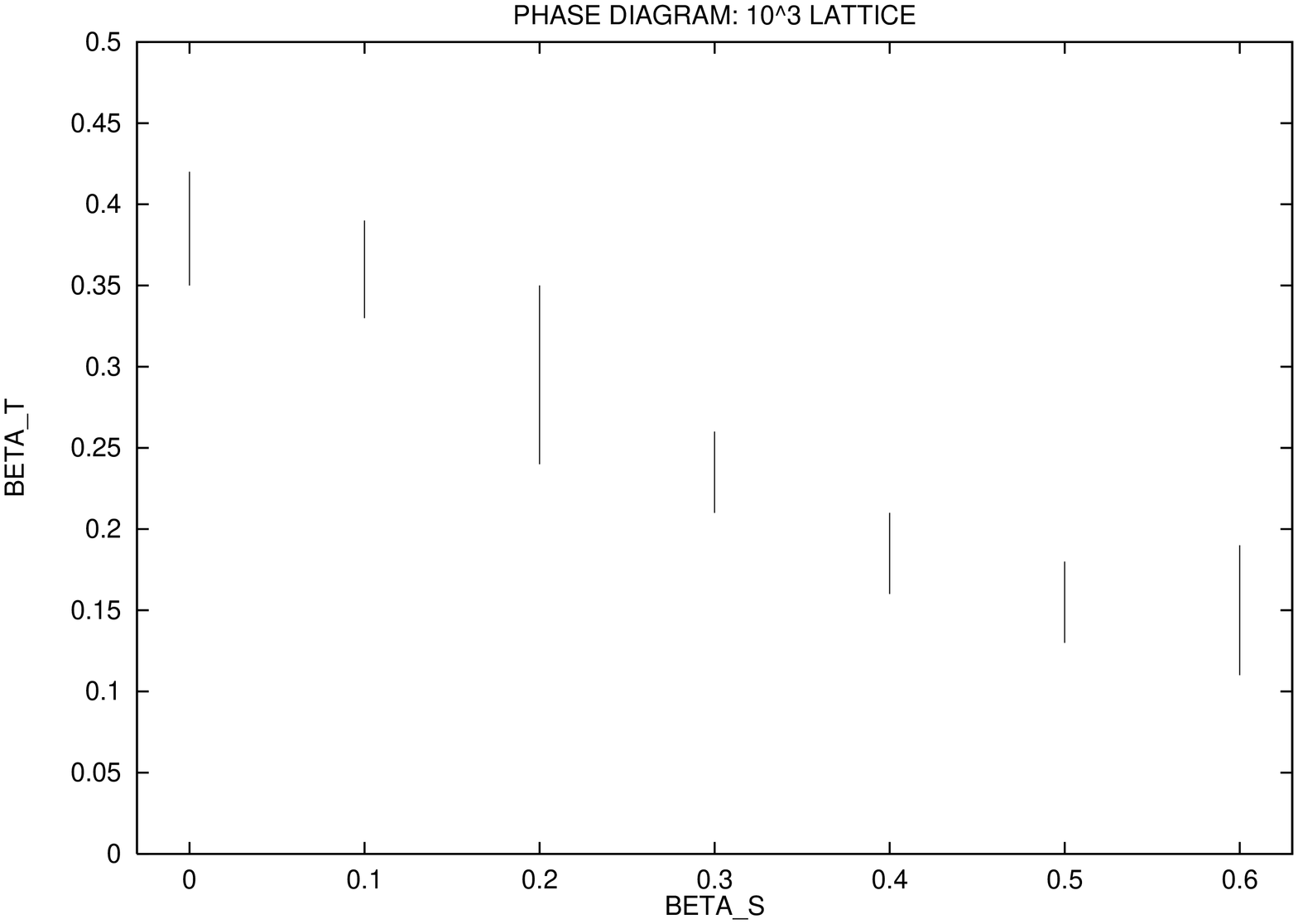,height=4cm,angle=-90}&
\psfig{figure=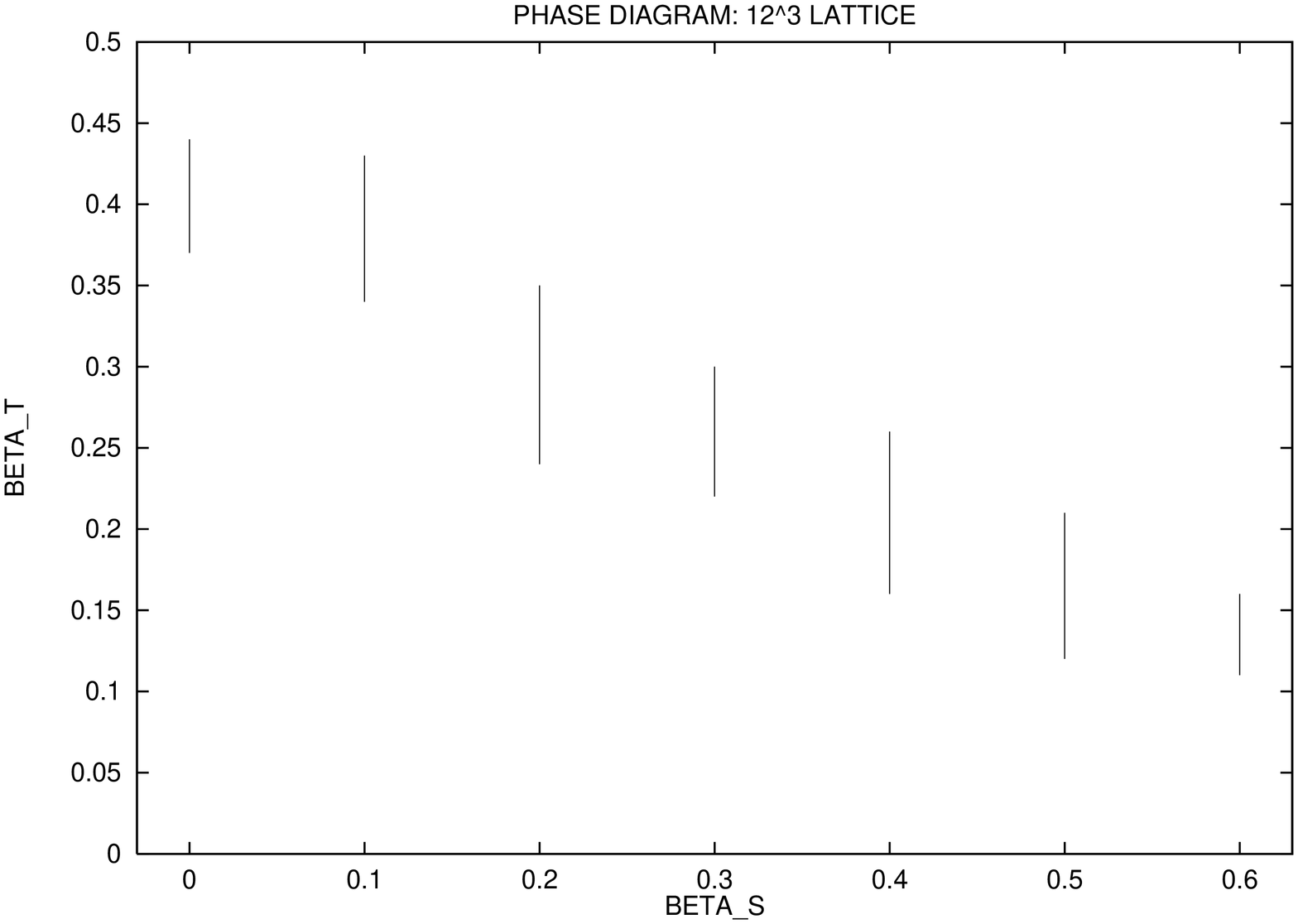,height=4cm,angle=-90}
\end{tabular}
\caption[pd]{Phase diagram for $\beta_S$ fixed at several values
for $6^3, 8^3, 10^3, 12^3$ lattices.}
\label{pd}
\end{figure}

For the first set of measurements we have fixed $\beta_S$ to several values,
let $\beta_T$ run and found the hysteresis loops
which have been formed. The results of
these measurements are displayed in figure \ref{pd}. The subfigures
correspond to the lattice volumes $6^3, 8^3, 10^3$ and $12^3$ respectively.
The line
segments indicate the extents of the hysteresis loops. We have proceeded
with steps of 0.005 for $\beta_T$ and performed 200 iterations at each
point. We have used plain Metropolis Monte Carlo as a simulation technique.
The phase transition line tends towards the horizontal axis for large
$\beta_S.$ It is not clear from such measurements
what happens for $\beta_S \rightarrow \infty,$
that is, whether the phase transition line meets the horizontal axis or
it ends at some point. However, it is known analytically
about the $\bt_T=0$ case that no phase transition should show up.
Conceivably the transition weakens and eventually becomes a
crossover before it meets the axis.

From the phase diagrams we may infer that the isotropic model will
have a phase transition at $\bt_S=\bt_T \approx 0.25.$ This value is
approximately the point where the $\bt_S=\bt_T$ line meets the phase
transition line.

After this first overview of the phase structure we will study some
of its characteristics in more detail. To this end we have performed long
runs, sticking to a particular point of the parameter space, that is
specific values for $\beta_S$ and $\beta_T$ and performing about three
million iterations. When the parameters are near a first order
phase transition we expect to see the eventual two-state signals.
As a by-product of this procedure, the critical points may be
determined with much greater accuracy than the one provided by
the hysteresis loop method.
We have found that, if we fix $\beta_S,$ the critical $\beta_T$ increases
slightly with the volume, while for the isotropic model the volume 
dependence is very small.




We now proceed with the presentation of the behaviour of the system
in the immediate vicinity of the phase transition. The
computer time evolution will be depicted for $8^3$ lattices.
We concentrated on three values of $\beta_S,$
namely 0.00, 0.20 and 0.60 and tried several values of $\beta_T$
near the critical point.

\begin{figure}
\begin{tabular}{cc}
\psfig{figure=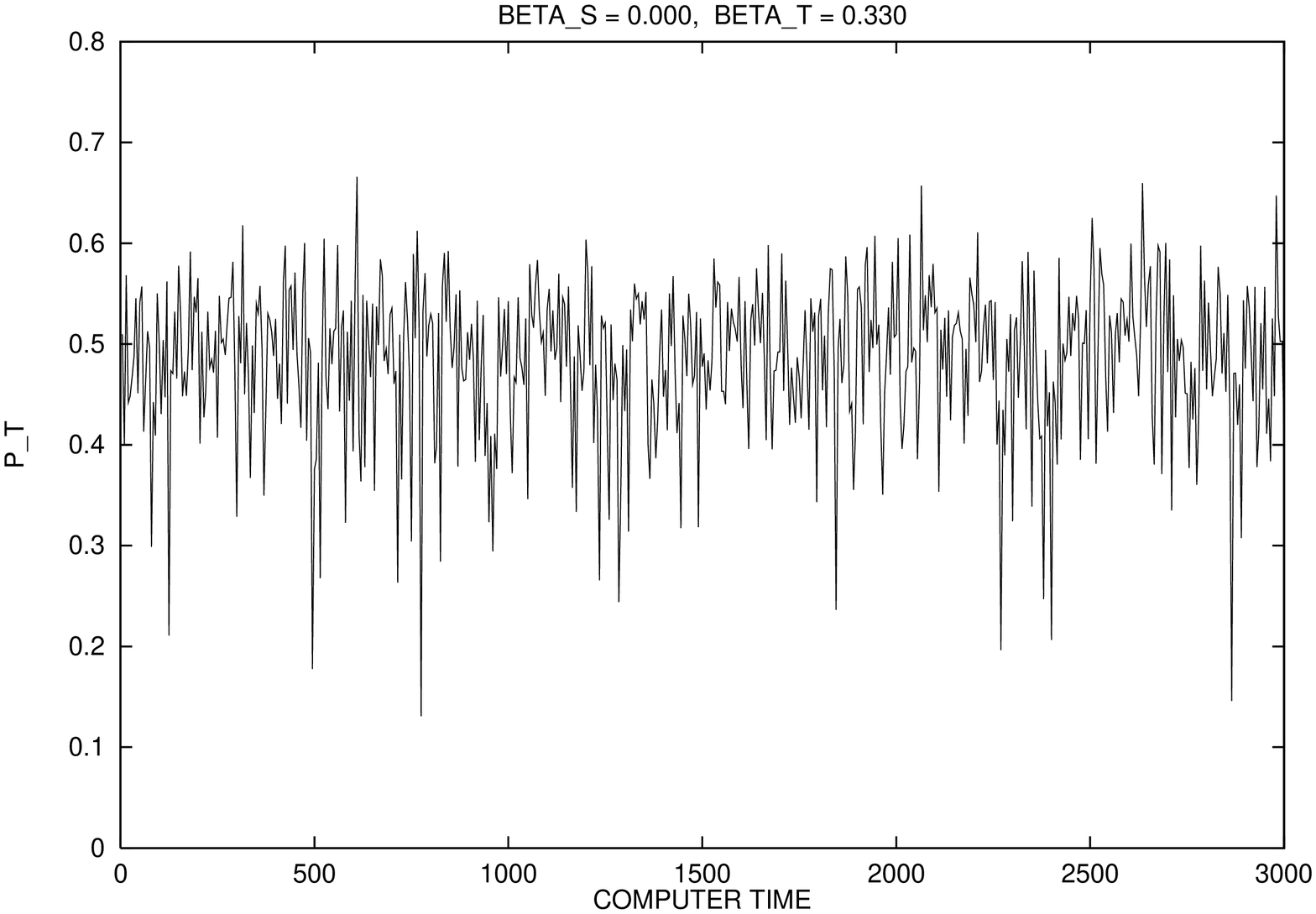,height=4cm,angle=-90}&
\psfig{figure=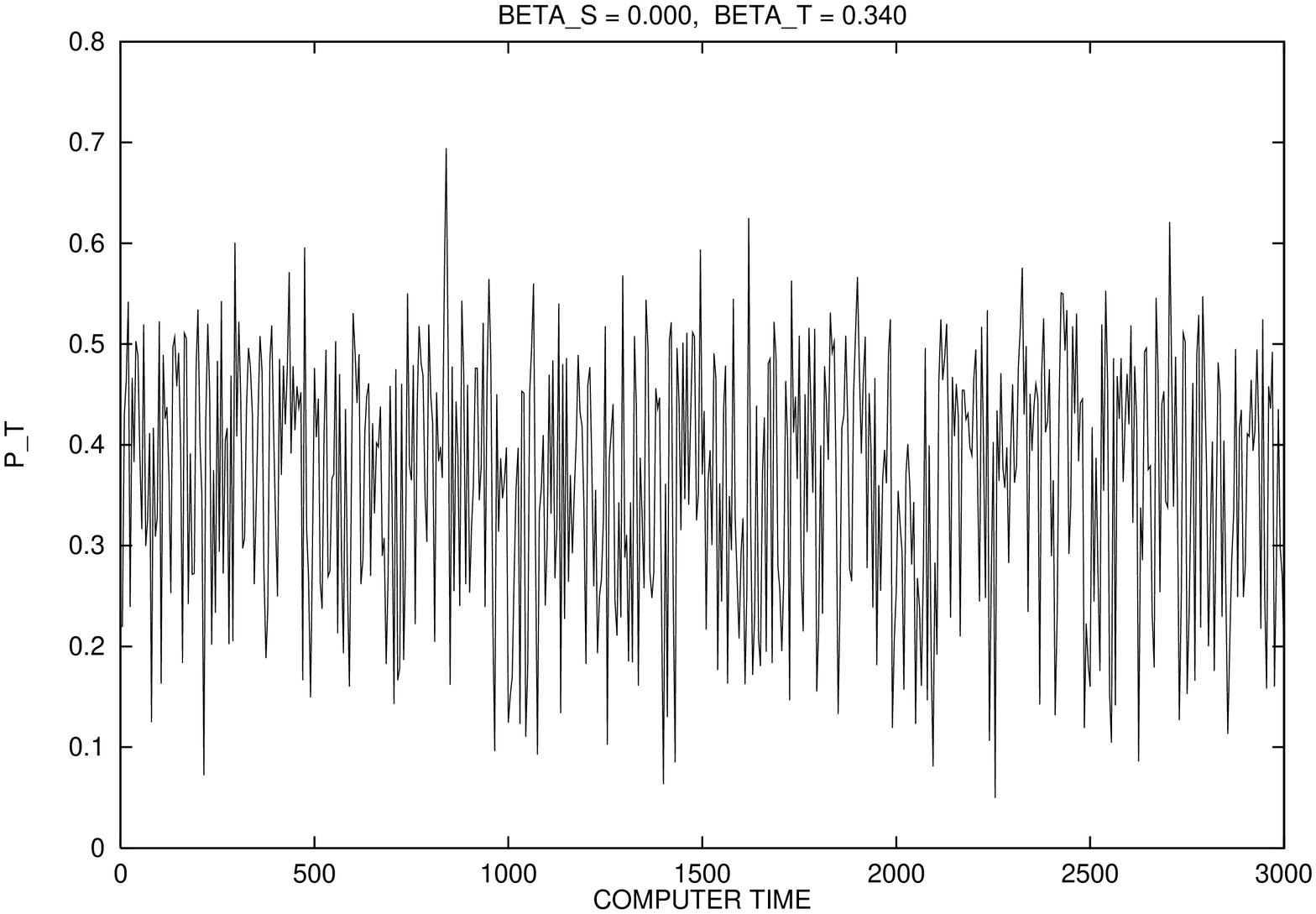,height=4cm,angle=-90}\\
\psfig{figure=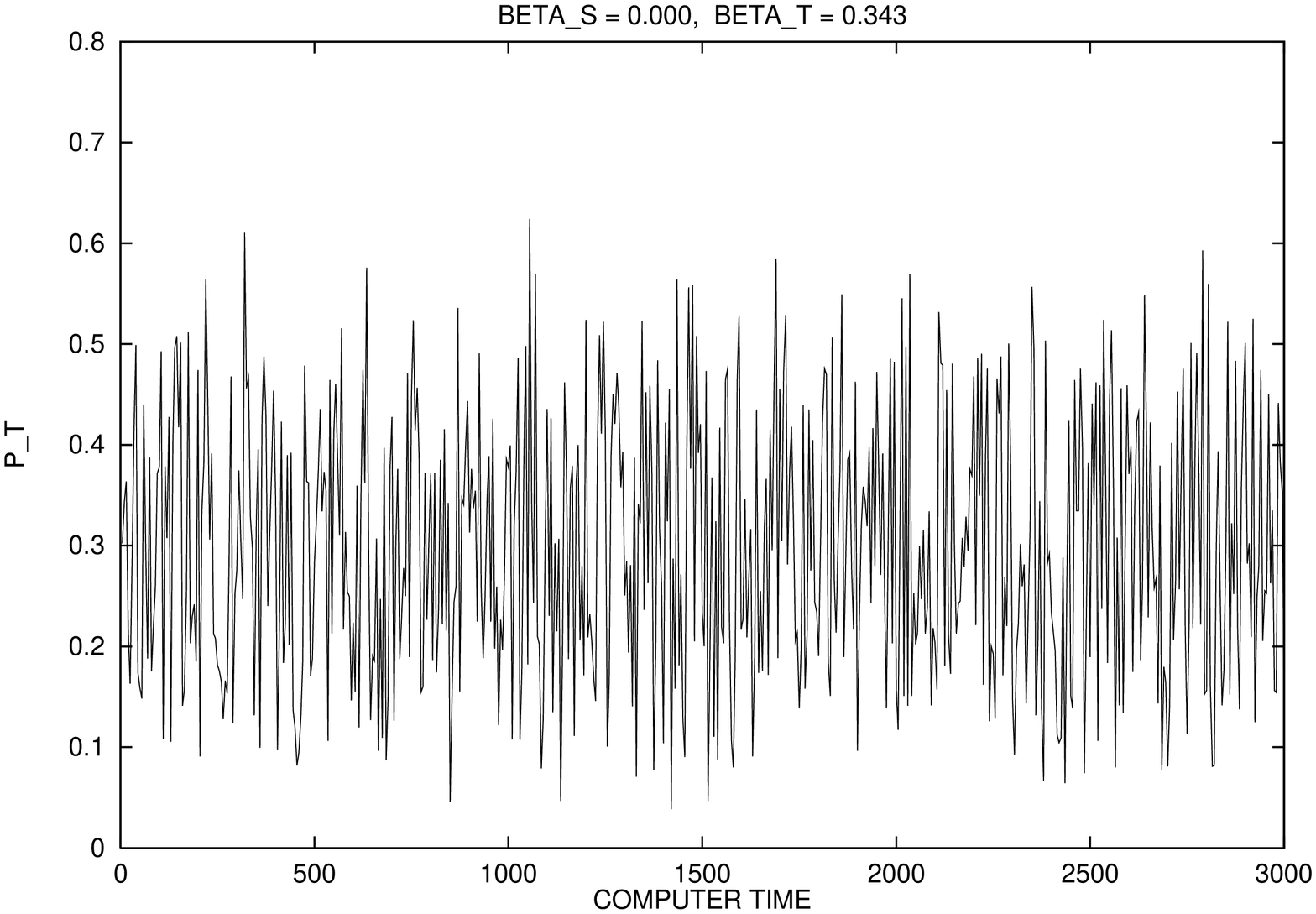,height=4cm,angle=-90}&
\psfig{figure=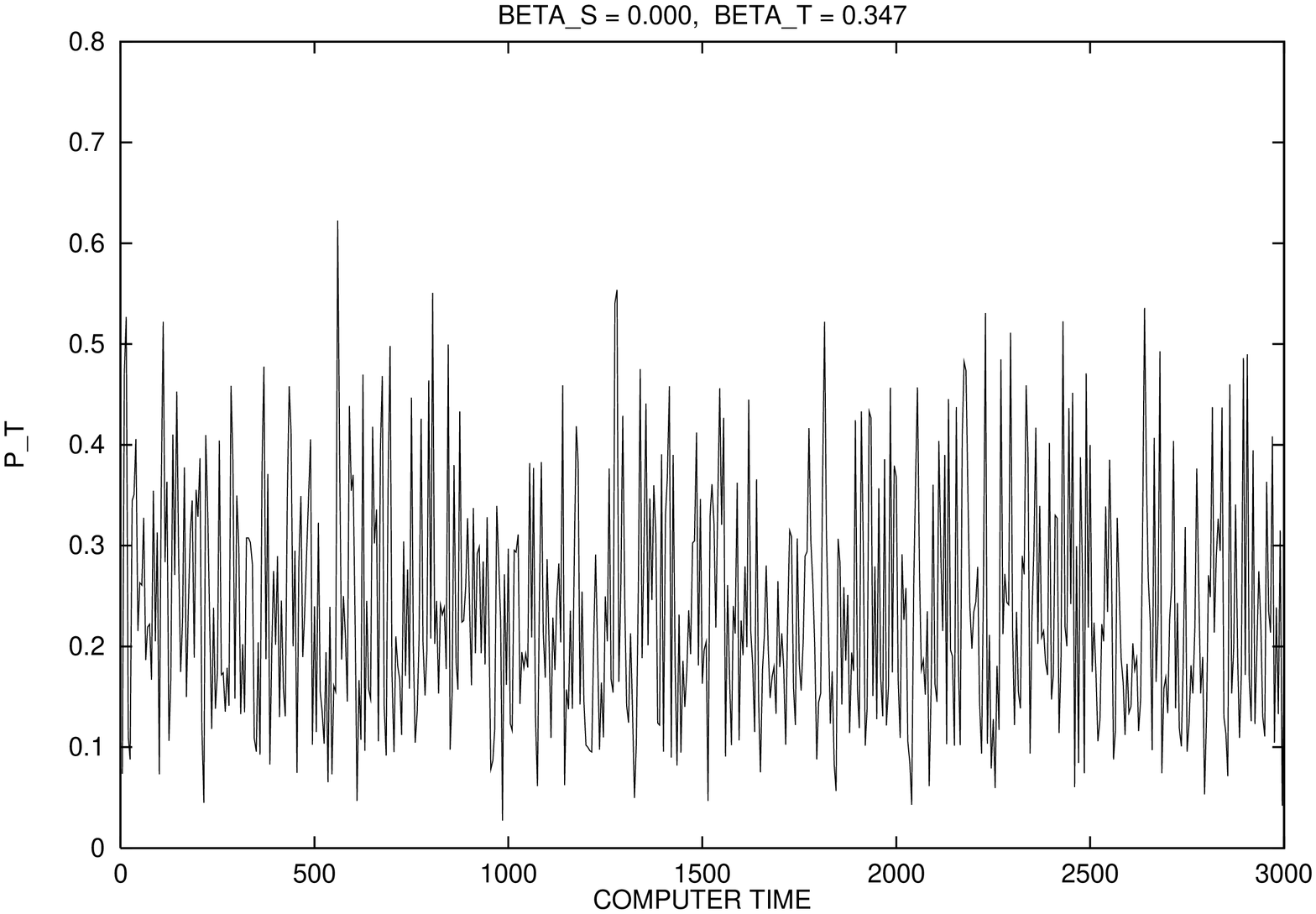,height=4cm,angle=-90}
\end{tabular}
\caption[lr000]{Time evolution of the transverse-like plaquette for a $8^3$
lattice at $\beta_S=0.000$ and $\beta_T=0.330, 0.340, 0.343, 0.347.$ }
\label{lr000}
\end{figure}

In figure \ref{lr000} we show the
time evolution of the transverse-like plaquette for a $8^3$ lattice for
$\beta_S=0.00$ and various values of $\beta_T$ around 0.34.
The system fluctuates violently and, as $\bt_T$ increases, it
gradually spends its time more and more in
the low $P_T$ region. This behaviour provides evidence
that the transition is of higher order.

\begin{figure}
\begin{tabular}{cc}
\psfig{figure=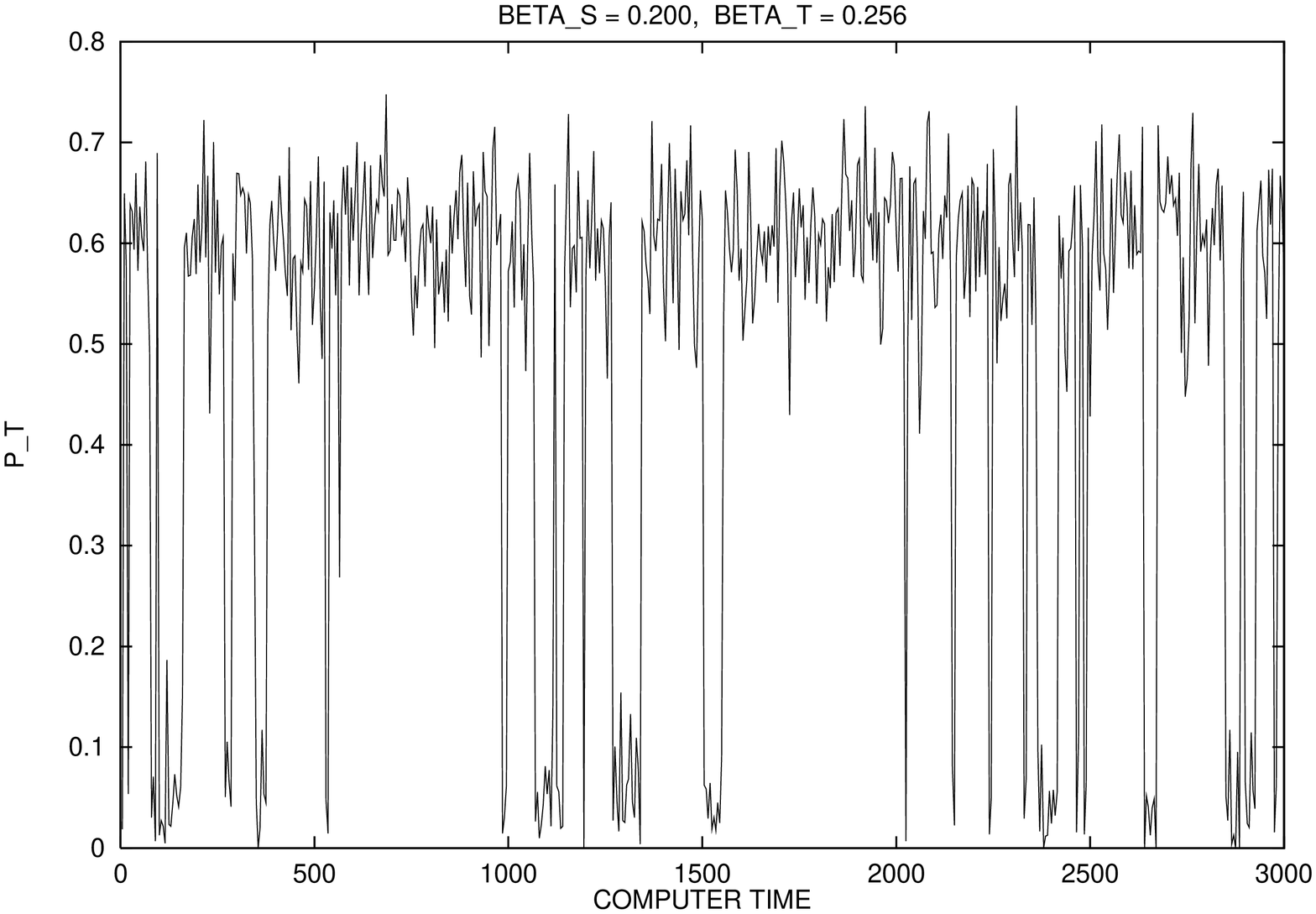,height=4cm,angle=-90}&
\psfig{figure=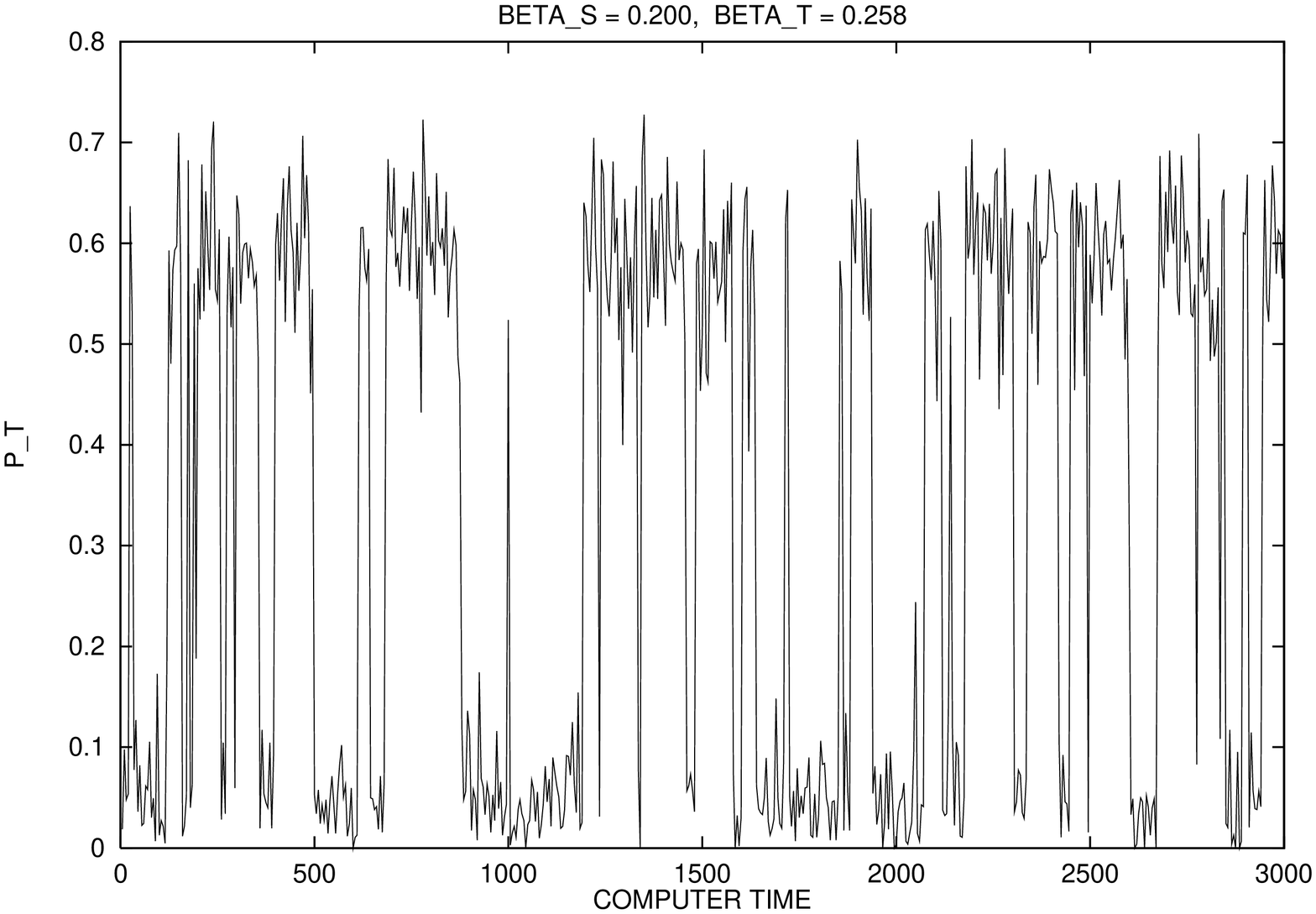,height=4cm,angle=-90}\\
\psfig{figure=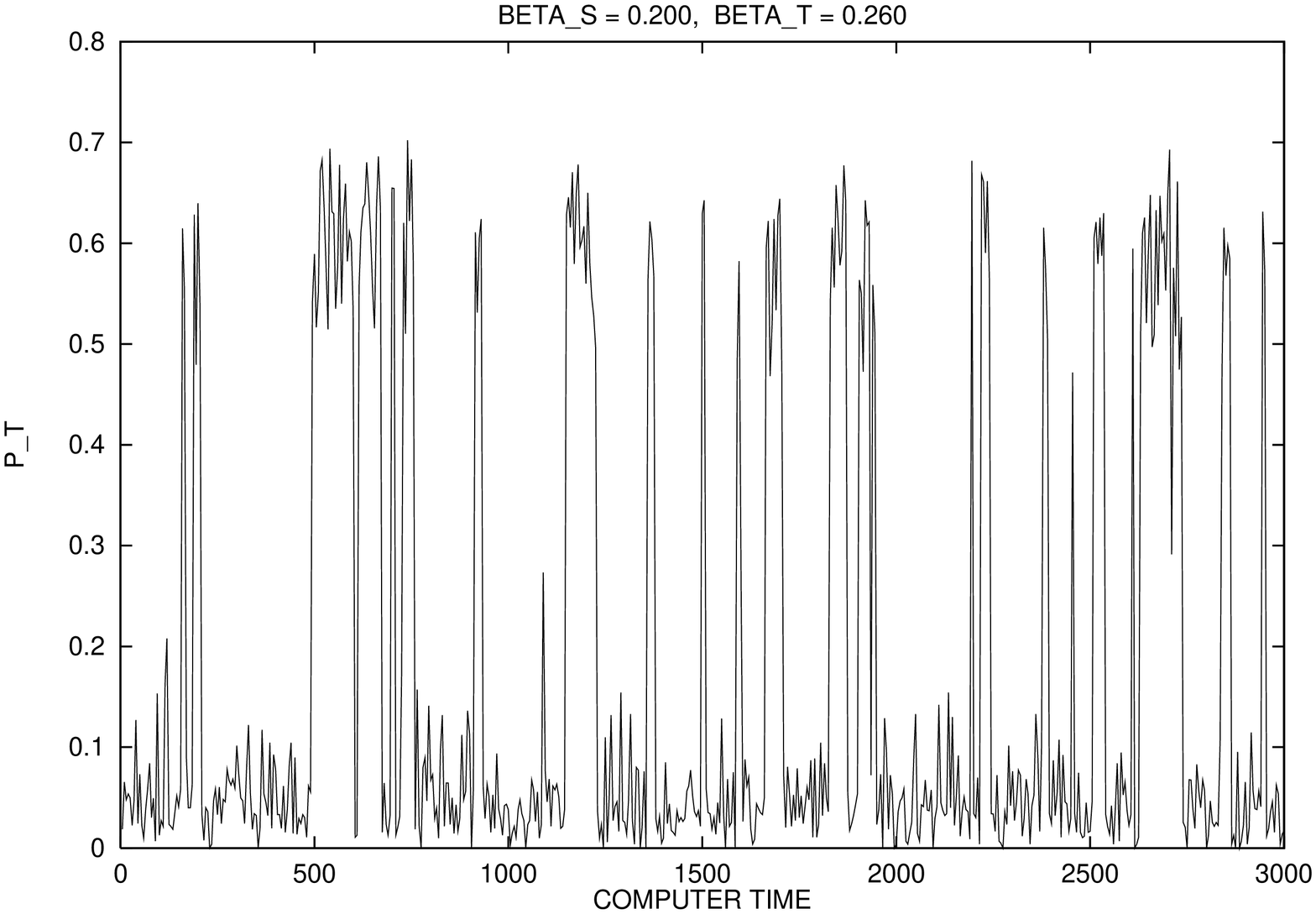,height=4cm,angle=-90}&
\psfig{figure=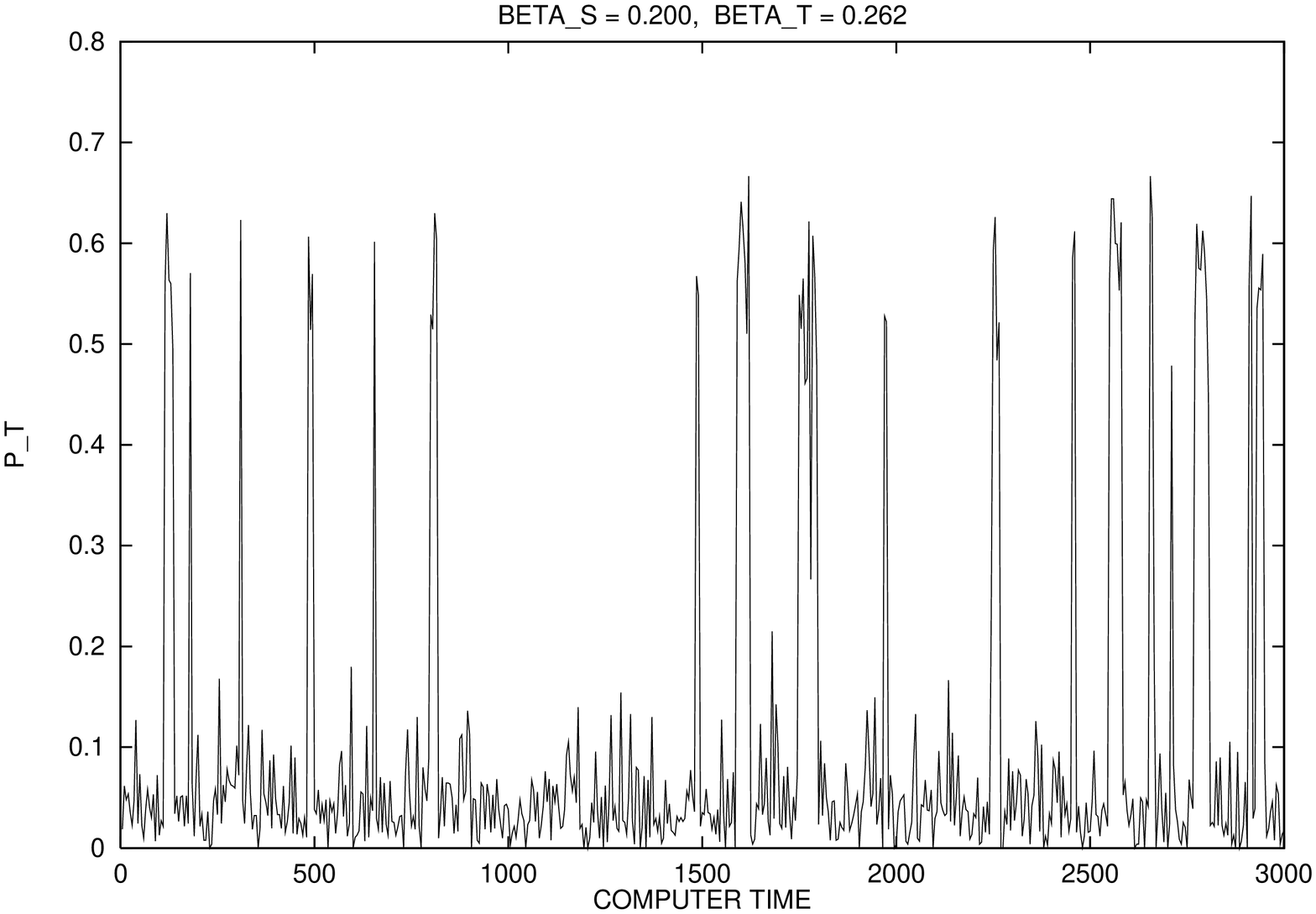,height=4cm,angle=-90}
\end{tabular}
\caption[lr020]{Time evolution of the transverse-like plaquette for a $8^3$
lattice at $\beta_S=0.200$ and $\beta_T=0.256, 0.258, 0.260, 0.262.$ }
\label{lr020}
\end{figure}

In figure \ref{lr020} we present the time evolution of
the transverse plaquette for $\beta_S=0.20$ in the phase transition region,
that is for $\beta_T$ around 0.26. We may clearly see the
oscillation of the mean values between the two metastable states and
the gradually increasing importance of the ``low $P_T$" metastable phase
with respect to the ``high $P_T$" one as $\bt_T$ increases.
More precisely, the system starts by spending
most of its ``time" in the state with large $P_T,$ but it gradually starts
visiting also the state with the small $P_T,$ until at some point it
spends most of its time in the small $P_T,$ as shown in the last subfigure.
A very similar behaviour shows up for $\beta_S=0.30.$

\begin{figure}
\begin{tabular}{cc}
\psfig{figure=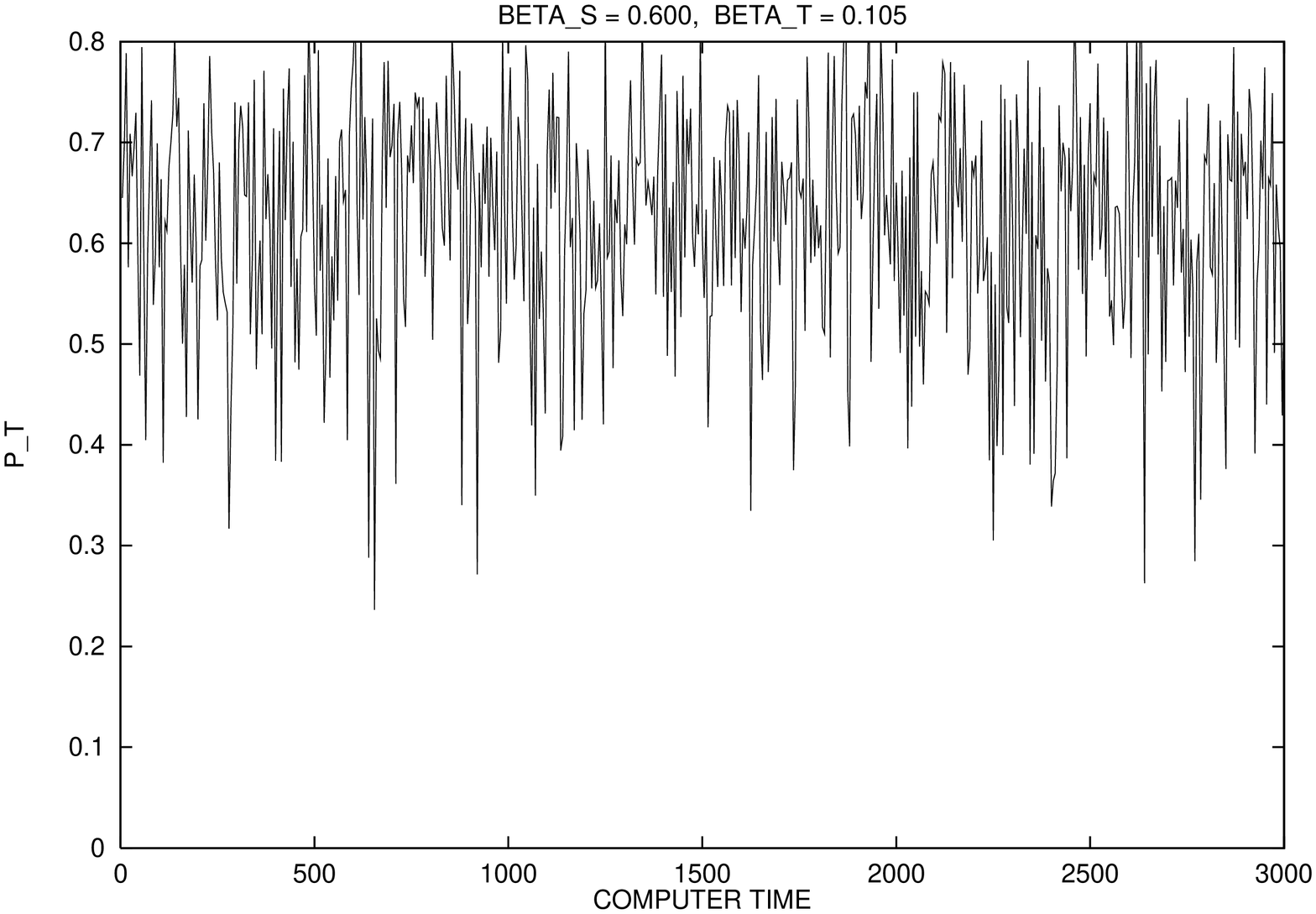,height=4cm,angle=-90}&
\psfig{figure=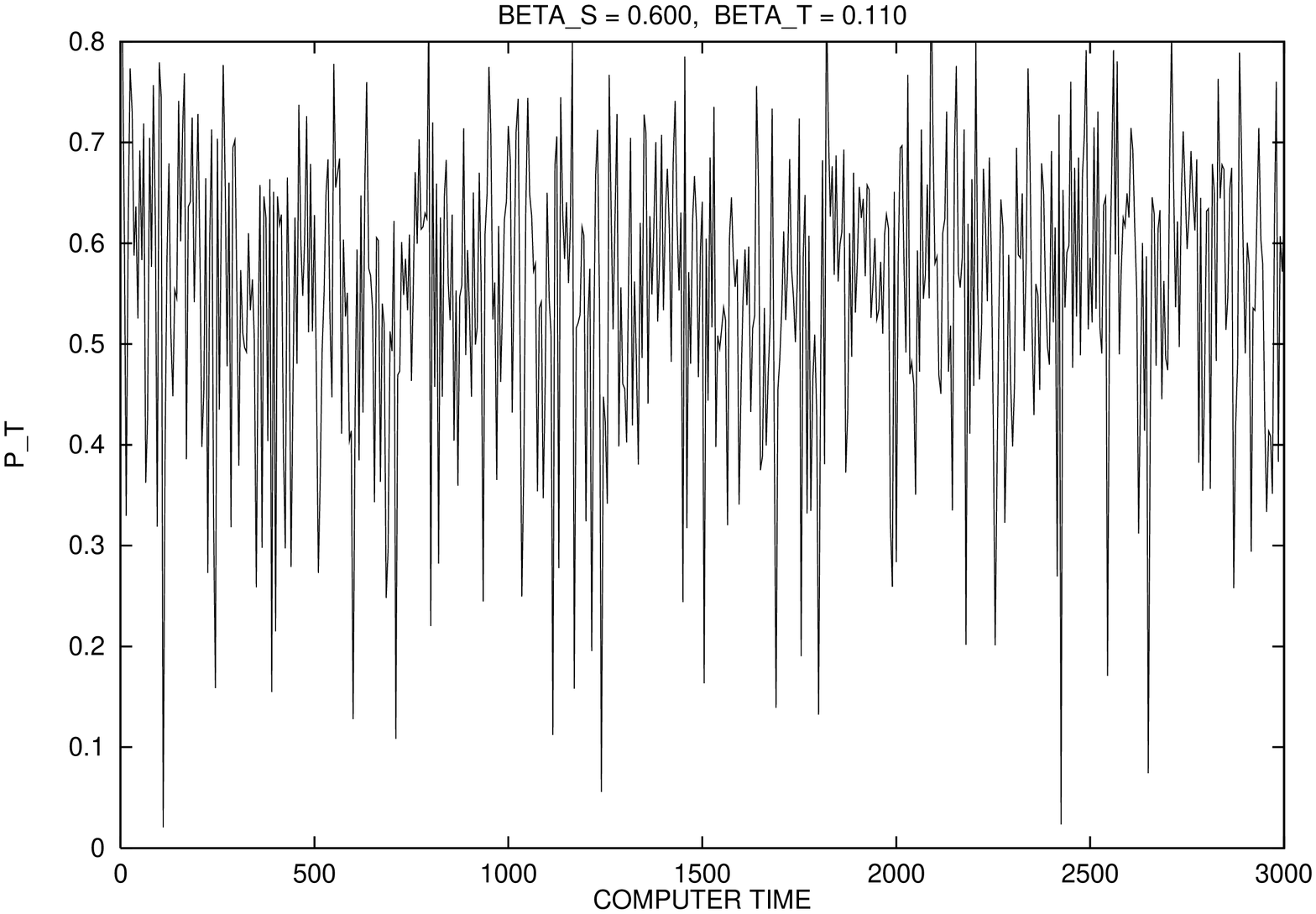,height=4cm,angle=-90}\\
\psfig{figure=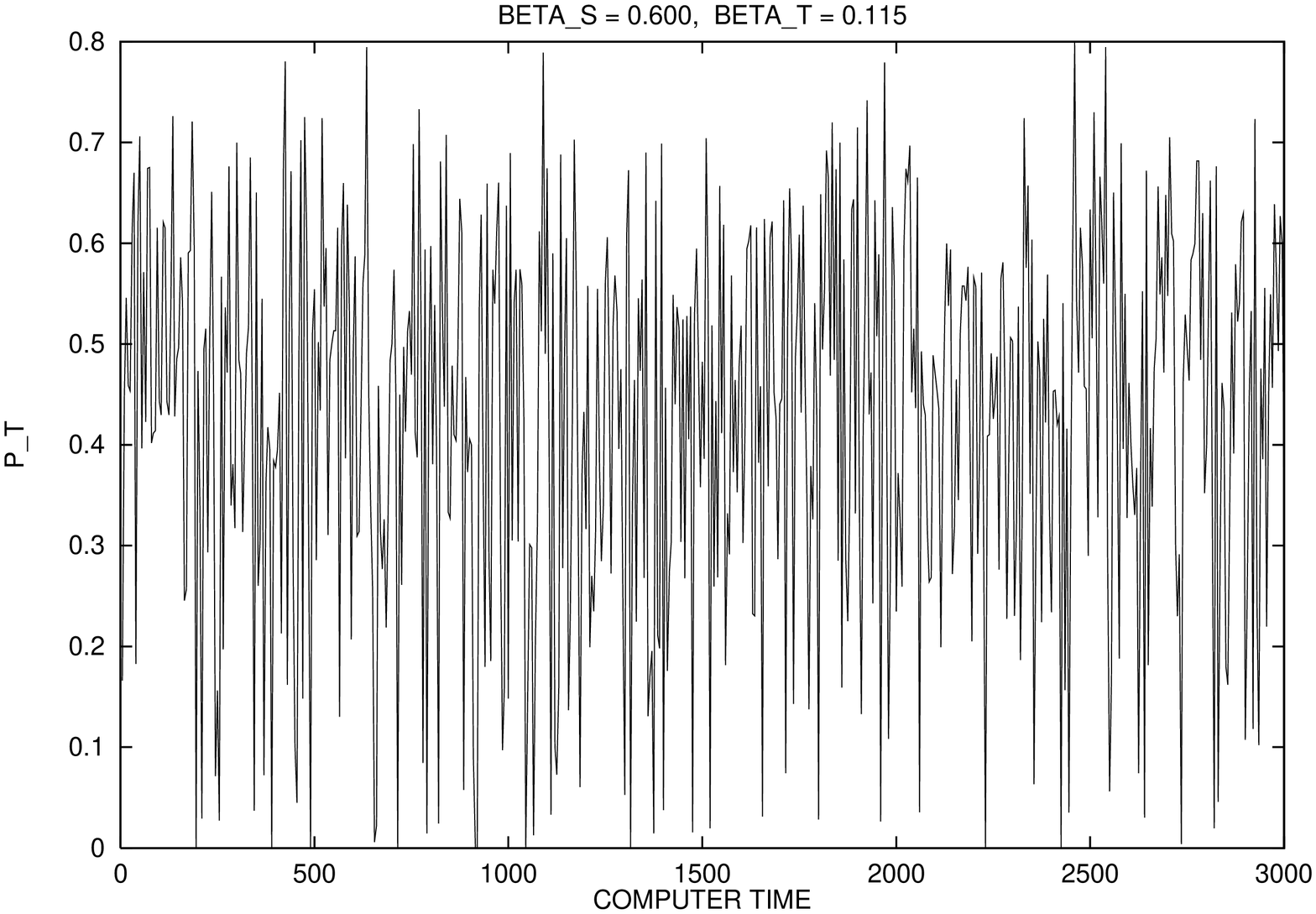,height=4cm,angle=-90}&
\psfig{figure=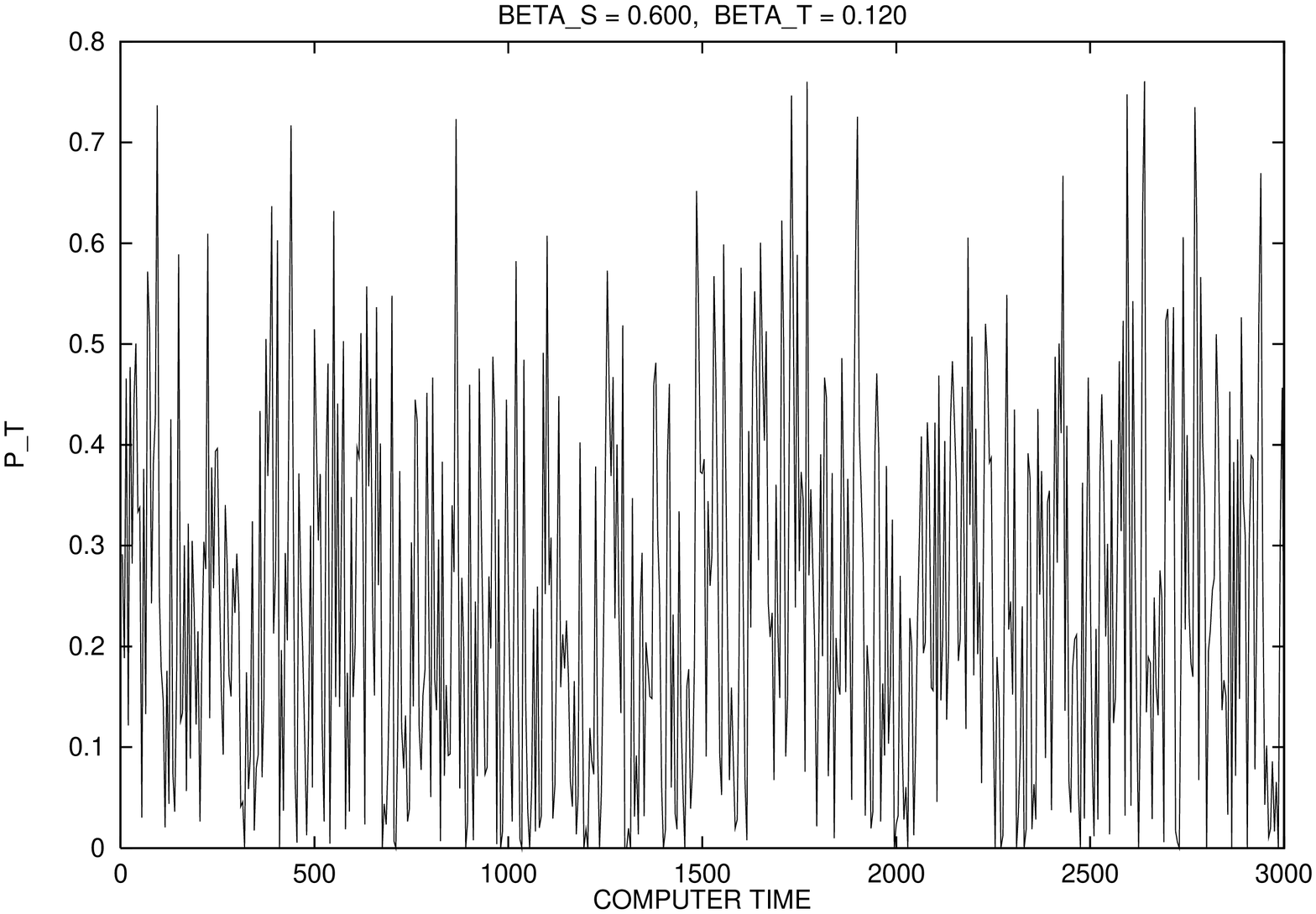,height=4cm,angle=-90}
\end{tabular}
\caption[lr060]{Time evolution of the transverse-like plaquette for a $8^3$
lattice at $\beta_S=0.600$ and $\beta_T=0.105, 0.110, 0.115, 0.120.$ }
\label{lr060}
\end{figure}

For $\beta_S=0.60$ the picture is strongly reminiscent of the
$\beta_S=0.00$ case.
Figure \ref{lr060} shows relatively long runs for several values of
$\bt_T$ around 0.11. The system performs large fluctuations again and
its mean value drifts toward small values of $P_T$ as $\beta_T$ increases.

It appears that the phase transition is first order around the value
$\beta_S \simeq 0.25,$ it remains strong for $\beta_S$ not very different
from this value, but it weakens substantially for too small or too
large values.

\begin{figure}
\begin{tabular}{cc}
\psfig{figure=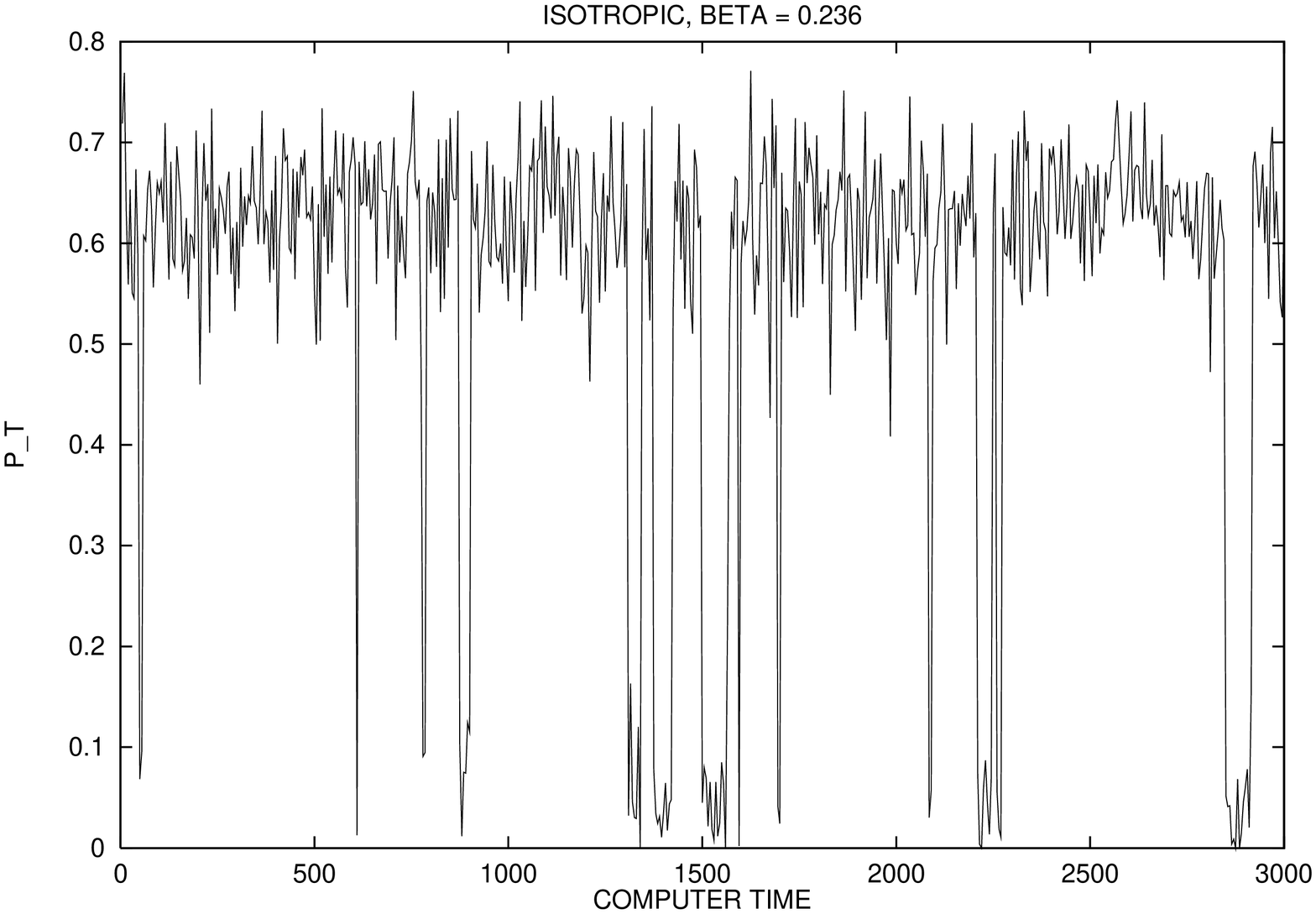,height=4cm,angle=-90}&
\psfig{figure=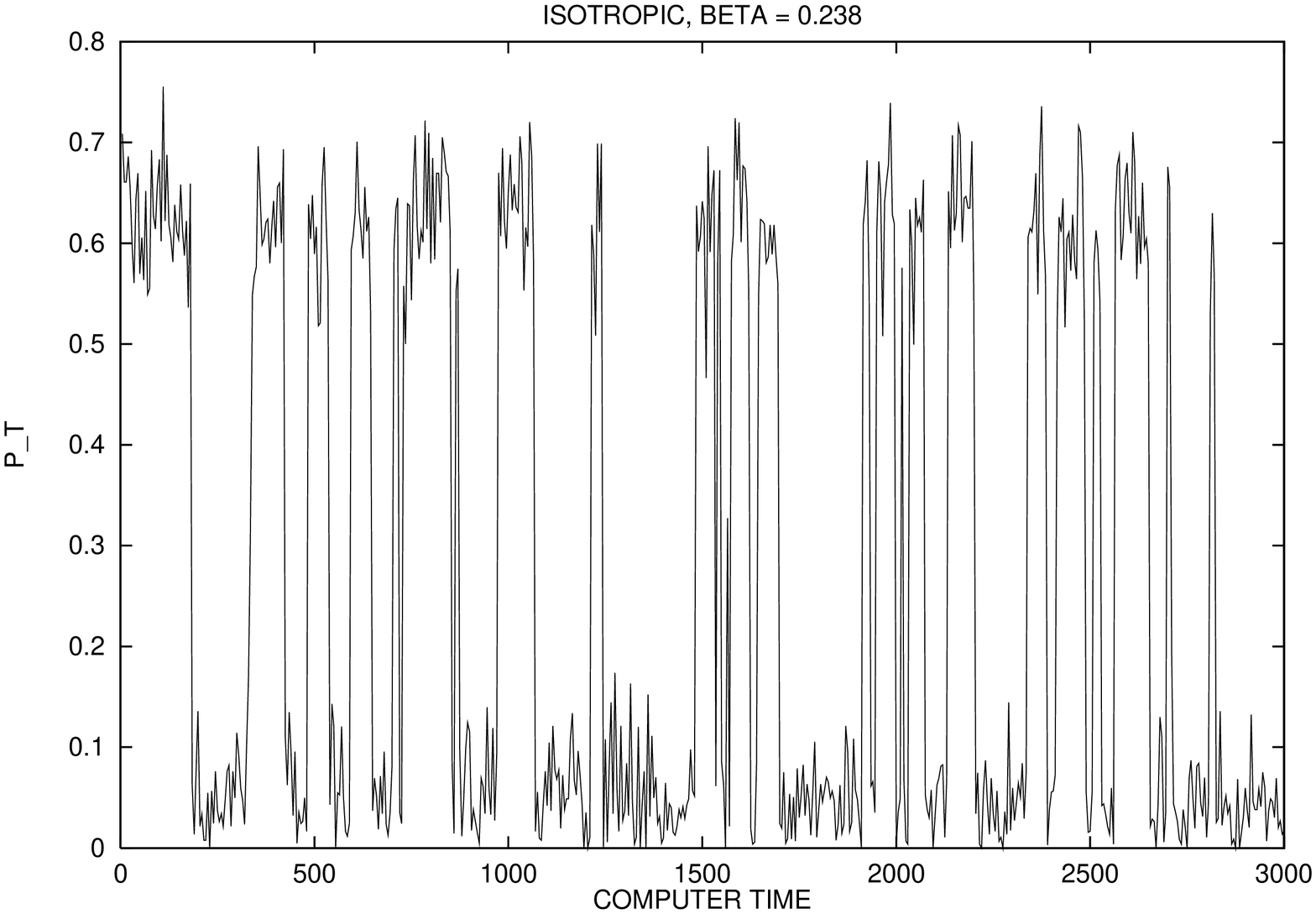,height=4cm,angle=-90}\\
\psfig{figure=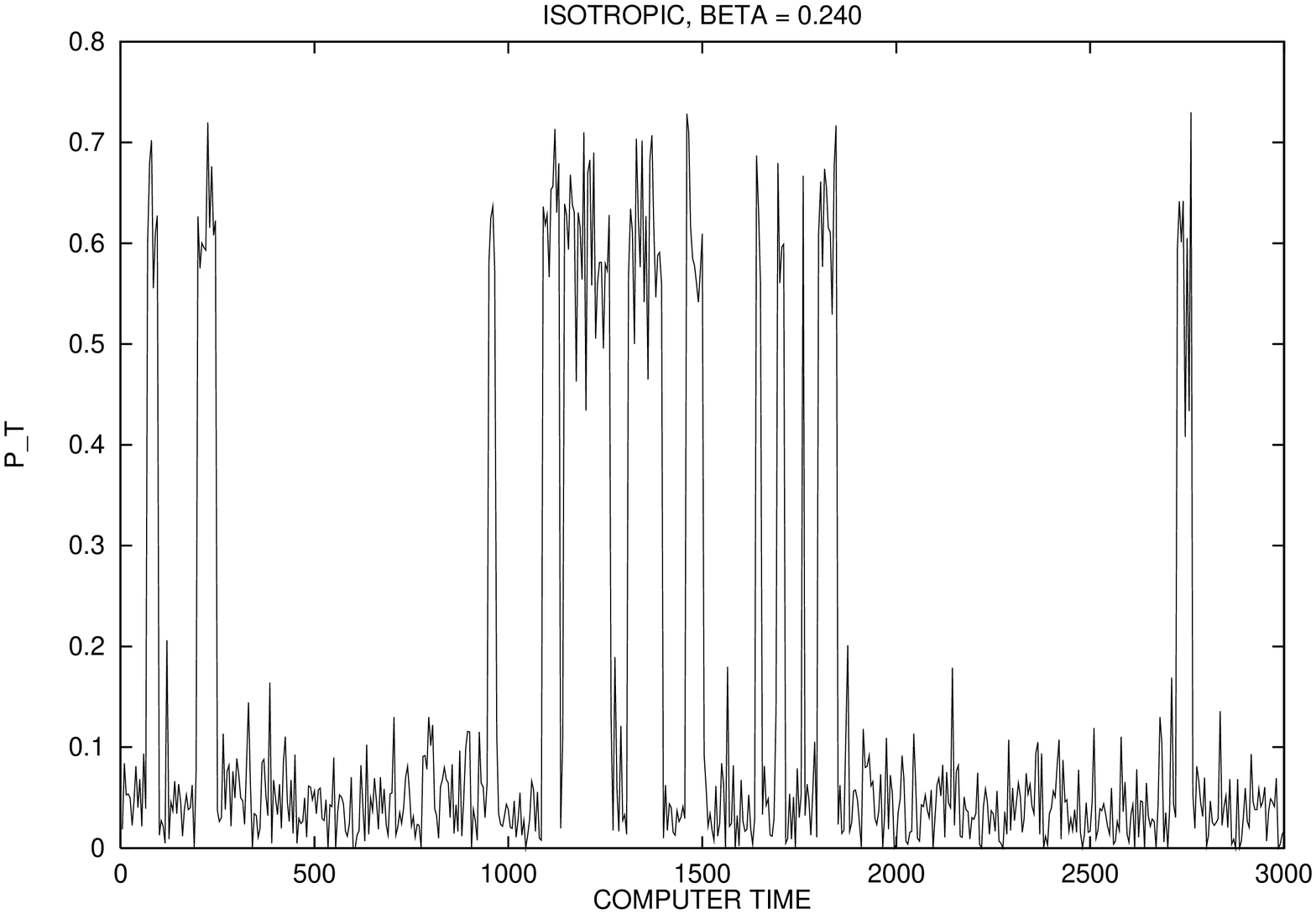,height=4cm,angle=-90}&
\psfig{figure=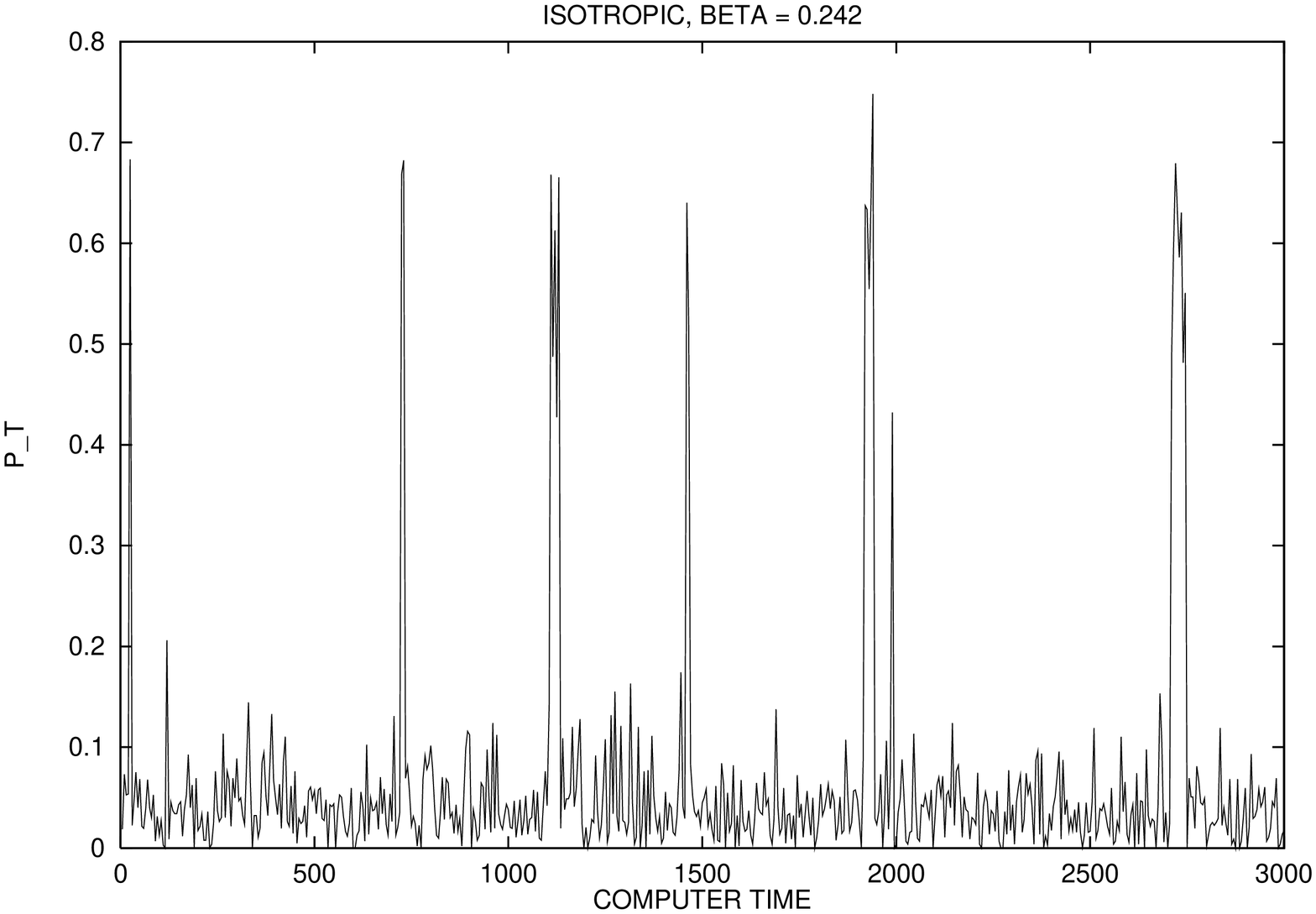,height=4cm,angle=-90}
\end{tabular}
\caption[lriso]{Long runs for the isotropic model
at $\bt_g=0.236, 0.238, 0.240$ and $0.242.$ }
\label{lriso}
\end{figure}

Finally, in figure \ref{lriso} we present the time evolution of the
plaquette for
the isotropic model on a $8^3$ lattice. For $\beta$ around 0.24 we observe
the phase transition and we may see the two metastable states.
The fluctuations of the system between the two metastable states
are very similar to the ones of the anisotropic model with
$\bt = 0.20.$ It appears that the phase transition is quite strong here,
in agreement with previously obtained results.

\begin{figure}
\centerline{\hbox{\psfig{figure=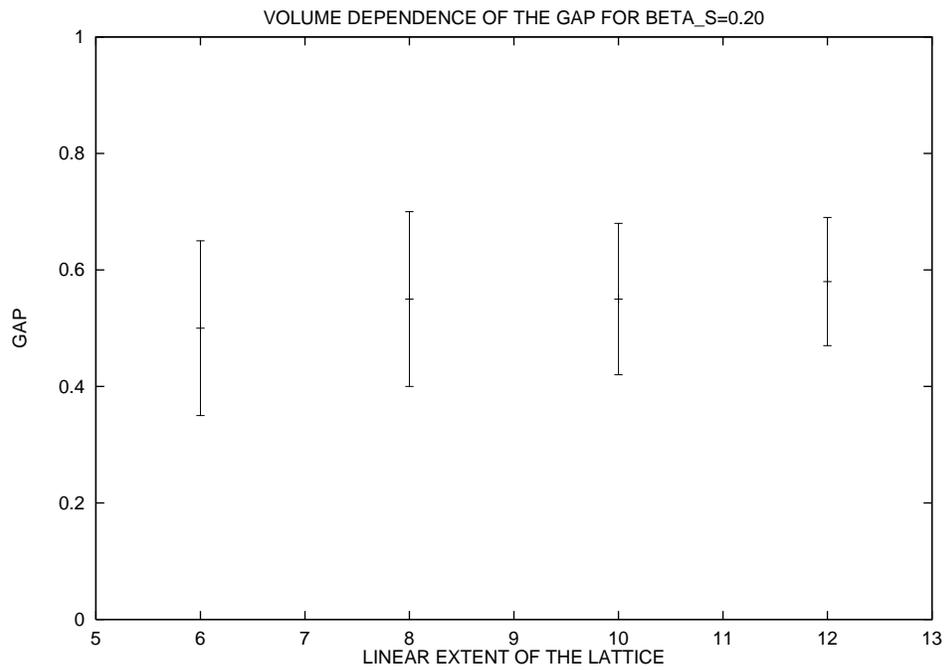,height=9cm,angle=-90}}}
\caption[gap]{The gap versus the volume.}
\label{gap}
\end{figure}

The final statement about the order of the phase transitions should come
from a study of the volume dependence of the susceptibilities and the
Binder cumulants. However, as one increases the volume, the system
sticks to either of the metastable states and one cannot really
observe the oscillation between the two states in a reasonable time.
The only exception occurs for rather small lattice volumes.
We have already presented the results for $8^3$ lattices, but it is
difficult to observe something similar for larger volumes.
However, one can easily see that if the susceptibility varies linearly
with the volume (which is the sign of a first order transition), the
gap between the two metastable states should be volume independent.
Thus, we may get an idea about the order of the phase transitions
by studying the volume dependence of the gap. If it is volume independent,
we have a first order transition. If it decreases with the volume, there is a
weaker phase transition (second or higher order). We find out that the
gap does not actually depend on the volume. Figure \ref{gap}
shows the results
for $\beta_S=0.20.$ Thus it appears that this phase transition is
of first order. The same picture also appears for
$\beta_S=0.30.$ This means that there are first order phase
transitions in a region around $0.25.$
For $\beta_S$ away from this region one cannot really define a gap
(compare figure \ref{lr060}).
It is clear, however, that the phase transition is weak in this regime.

\section{Acknowledgements} One of the authors G.S. was supported in part
by the EEC Grant HPRN CT 199900161.  G.K. acknowledges the
support from EEC Grant no. ERBFMRX CT 970122 and he would like to
thank P.Dimopoulos for useful discussions.

\vfill
\end{document}